# Ultra-high bandwidth fiber-optic data transmission with a single chip source

**David J. Moss**
*Optical Sciences Centre, Swinburne University of Technology, Hawthorn, VIC 3122, Australia*

**Keynote Words**— Optical data transmission, fiber optics, microwave photonics, micro-ring resonators, optical micro-combs.

*Abstract*—We report world record high data transmission over standard optical fiber from a single optical source. We achieve a line rate of 44.2 Terabits per second (Tb/s) employing only the C-band at 1550nm, resulting in a spectral efficiency of 10.4 bits/s/Hz. We use a new and powerful class of micro-comb called soliton crystals that exhibit robust operation and stable generation as well as a high intrinsic efficiency that, together with an extremely low spacing of 48.9 GHz enables a very high coherent data modulation format of 64 QAM. We achieve error free transmission across 75 km of standard optical fiber in the lab and over a field trial with a metropolitan optical fiber network. This work demonstrates the ability of optical micro-combs to exceed other approaches in performance for the most demanding practical optical communications applications.

## I. INTRODUCTION

Kerr micro-combs [1-4] offer the full potential of their bulk counterparts [5,6] but in an integrated footprint, since they generate optical frequency combs in integrated micro-cavity resonators. The realization of soliton temporal states called dissipative Kerr solitons (DKSs) [7-11] opened up a new method of mode-locking micro-combs that has given rise to major breakthroughs in spectroscopy [12,13], microwave and RF photonics [14], optical frequency synthesis [15], optical ranging including LIDAR [16, 17], quantum photonic sources [18-21], metrology [22, 23] and much more. One of their most promising applications has been in the area of optical fibre data communications where they have formed the basis of massively parallel multiplexed ultrahigh capacity optical data transmission [4, 24 - 26]. In this paper, by employing a powerful new type of micro-comb based on soliton crystals [11], we report a world record speed of data transmission across standard optical fibre from any single optical source. We achieve a line rate of 44.2 Terabits/s (Tb/s) utilizing only the C-band, and achieve a very high spectral efficiency of 10.4 bits/s/Hz. Spectral efficiency is critically important since it directly governs how much total bandwidth can be realized in a system. Soliton crystals display very stable and robust operation and generation as well as a very high intrinsic conversion efficiency that, all taken together with the low soliton micro-comb FSR of 48.9 GHz, enabled us to use a high coherent modulation data format of 64 quadrature amplitude modulation (QAM). We demonstrate error free data transmission across 75 km of standard optical fibre in our lab and in a field trial in an installed metropolitan area optical fibre testbed network in the Melbourne region. Our results were underpinned by the ability of soliton crystals to operate without stabilization or feedback control, but only open loop systems. This significantly reduced the amount of instrumentation required. Our work directly proves the capability of optical Kerr microcombs to out-perform other approaches for optical communications systems.

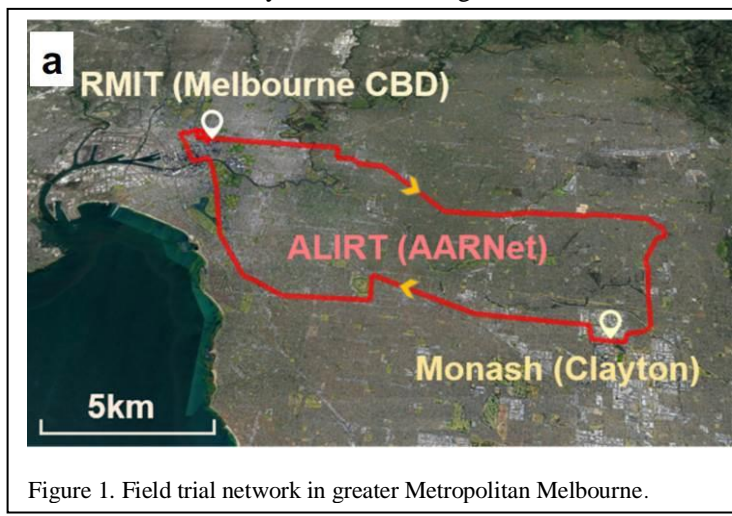

Figure 1. Field trial network in greater Metropolitan Melbourne.

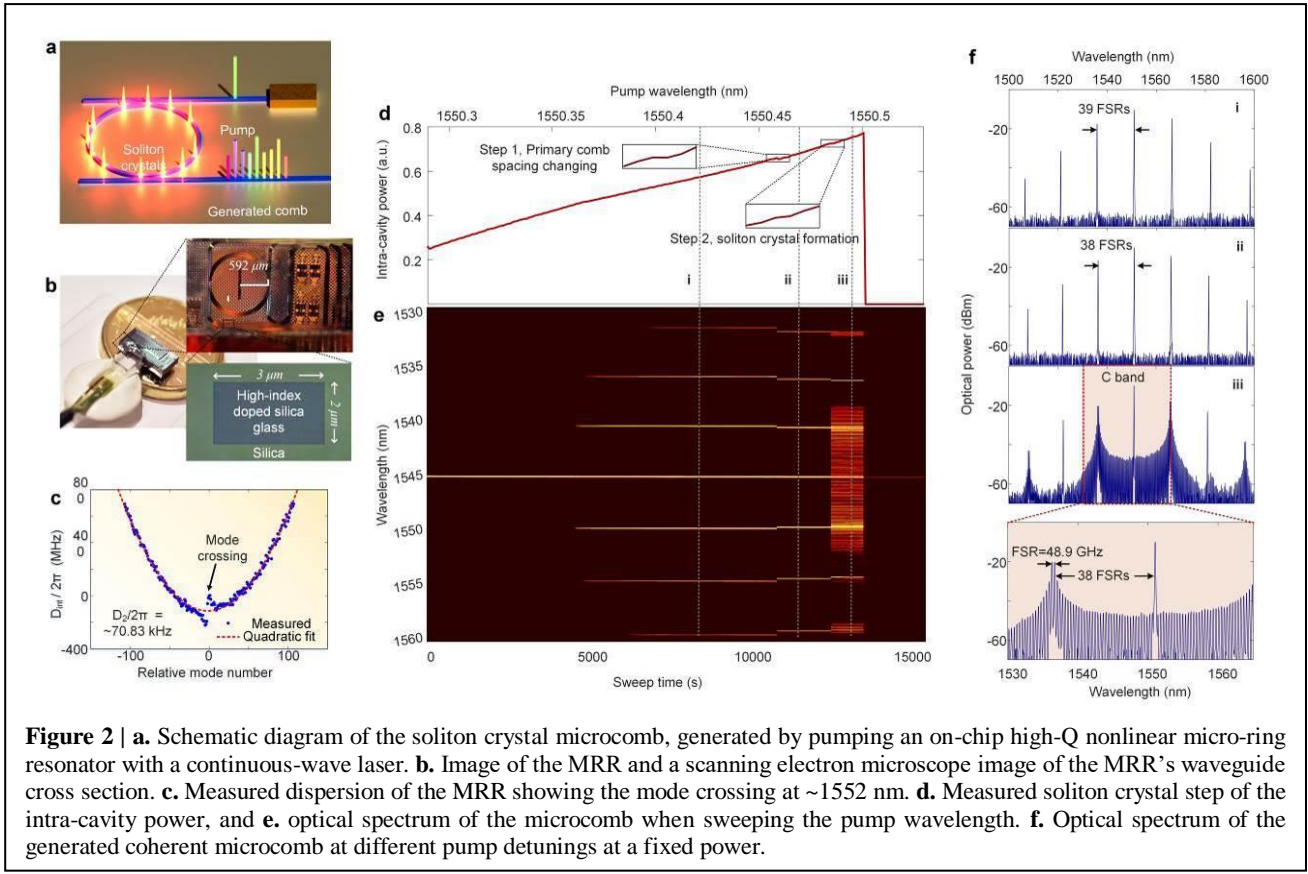

**Figure 2 | a.** Schematic diagram of the soliton crystal microcomb, generated by pumping an on-chip high-Q nonlinear micro-ring resonator with a continuous-wave laser. **b.** Image of the MRR and a scanning electron microscope image of the MRR's waveguide cross section. **c.** Measured dispersion of the MRR showing the mode crossing at ~1552 nm. **d.** Measured soliton crystal step of the intra-cavity power, and **e.** optical spectrum of the microcomb when sweeping the pump wavelength. **f.** Optical spectrum of the generated coherent microcomb at different pump detunings at a fixed power.

Currently, 100's of Tb/s are transmitted every instant across the world's fibre optic networks and the global bandwidth is growing at a rate of 25% /yr [27]. Ultrahigh capacity data links that use parallel massive wavelength division multiplexing (WDM) systems combined with coherent advanced modulation formats [28], are critical to meet this demand. Space-division multiplexing (SDM) is another emerging approach where multiple signals are transmitted either over multiple core or multiple mode fibre, or both [29]. In parallel with all of this, there is a growing movement towards very short links but still with very high capacity, particularly for data centres. Even just ten years ago, long haul networks such as undersea links spanning thousands of kilometres, used to dominate the global infrastructure, but nowadays the demand has dramatically shifted towards smaller scale applications including data centres as well as metropolitan area networks (10s to 100s of kilometres in size). These trends demand highly compact, energy efficient and low-cost devices. Photonic integrated circuits are the only approach that can address these needs, where the optical source is absolutely key to each link, and therefore has the greatest need to meet these requirements. The ability to generate all wavelengths on a single chip that is both integrated and compact, replacing many lasers, will yield the highest benefits [30-32].

Kerr optical microcombs have attracted a great deal of interest and one of their main applications has been in this area. They have successfully been used as optical sources for ultra-high bandwidth optical fiber data transmission [24 - 26]. A key factor has been achieving the capacity to modelock all of the microcomb lines, and this has been characterized by the discovery of new states of temporal optical soliton oscillation that include feedback-stabilized Kerr combs [25], dark solitons [32] and dissipative Kerr solitons (DKS) [24]. The last one (DKS) has achieved the greatest success, being the basis of extremely high data transmission rates across the full C and L telecom bands, at a rate of 30 Tb/s using only a single source, and 55 Tb/s using two microcombs [24]. Despite this, micro-combs need to be even more stable and simpler and robust in both operation and generation, in order to meet the demands of real-world installed fibreoptic systems [26. 28 - 32]. They particularly must work without the need for complicated stabilization feedback, preferably in uncomplicated open-loop fashion and without the need for complicated pumping schemes that DKS states need in order to be generated. Furthermore, the conversion efficiency from pump to comb lines must be much higher and their threshold pump power much lower. Systems that use microcombs also must achieve a much higher spectral efficiency (SE) since to date they have only achieved about ¼ of the theoretical maximum. Spectral efficiency is an absolutely key and fundamental parameter that limits the total data capacity of systems [28, 29].

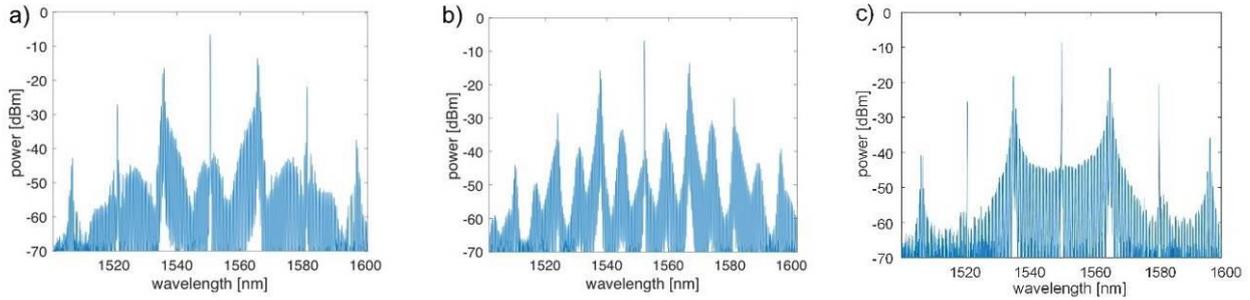

**Figure 3.** Alternate comb generation states and comb stability measurements. a) Measured spectrum of an alternate soliton crystal generation state achieved through manual wavelength tuning of the pump laser. b) Captured spectrum of a modulation instability ('chaotic') state, also achieved via manual wavelength tuning. c) Final complete soliton crystal spectrum used in the transmission experiments.

This work reports a world-record high bandwidth for optical fibre data transmission using standard single mode fibre together with a single optical source. Our use of soliton crystals [11, 26], based on CMOS - compatible chips [2, 3, 33 - 50], enabled us to reach a transmission data rate of 44.2 Tb/s using only a single chip – an increase of almost 50% [24, 25]. More importantly, we report an improvement, by a factor of 3.7, in the SE, achieving 10.4 bits / s / Hz -  a record high value for microcombs. We do this through the use of a high coherent modulation format of 64 QAM, together with a microcomb that has a very low spacing, or FSR, of 48.9 GHz. We only use the telecom C-band, leaving room for significant expansion in our capacity. We report experiments in the lab with 75 km of fibre as well as over an installed metropolitan optical fibre network (Figure 1). These results were made possible because of the highly and stable and robust generation and operation of the soliton crystals, together with their very high natural efficiency.

Soliton crystal oscillation states in micro-resonators that have a crystalline type of profile along the resonator path, forming in the angular domain of tightly packed self-localized pulses within micro-ring resonators [11]. They can occur in integrated ring resonators that have a higher order mode crossing. Further, they do not need the dynamic and very complicated pumping schemes or elaborate stabilization that self-localised DKS states need [51]. The basis of their stable behaviour originates from the fact that their intra-cavity power is dramatically higher than DKS states. In fact, it is very similar to the power levels of the chaotic temporal states [11, 52]. As a result, there is a very small difference in power levels in the cavity when the soliton crystal states are created out of chaos, and so there is no change in the resonant frequency. It is this self-induced frequency detuning arising from thermal instability due to the soliton step that renders pumping of DKS states, for example, challenging [53]. The combined effect of natural stability and robust and simple manual generation and the overall efficiency of soliton crystals that makes them extremely attractive for very high bandwidth data transmission.  We note that this experiment provides the first communication system demonstrated using a soliton crystal state, with previous optical systems demonstrations focusing on applications in microwave photonics [56].

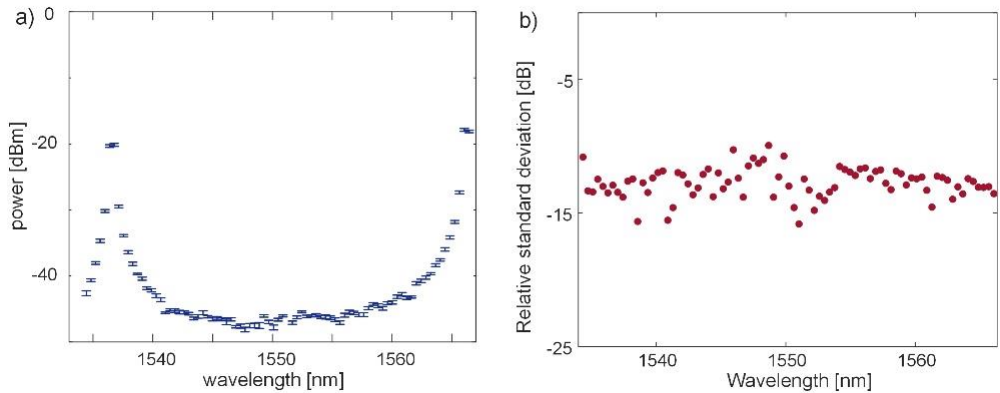

**Figure 4. Power stability measurements**. a) zoomed in soliton spectrum of the 80 channels selected over the C-band for the experiments, along with the power standard deviation (error bars) of the comb lines over 66 hours (with traces captured at 15 minute intervals), and b) relative power standard deviation (in dB).

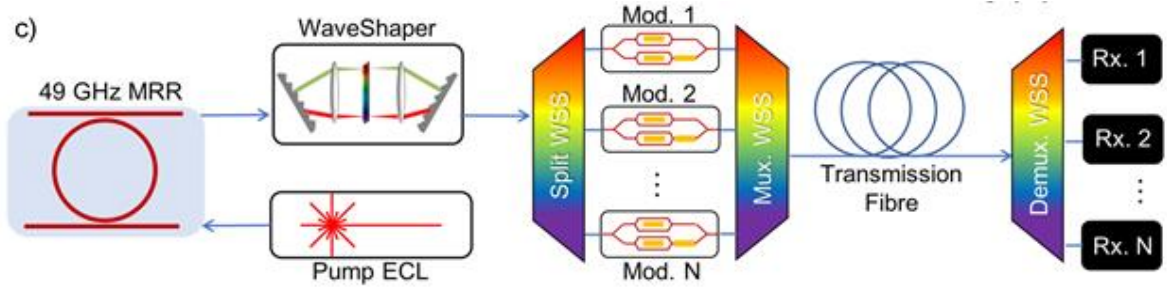

**Figure 5. Soliton crystal micro-comb communications experiment.** A CW laser, amplified to 1.8W, pumped a 48.9 GHz FSR micro-ring resonator, producing a micro-comb from a soliton crystal oscillation state. The comb was flattened and optically demultiplexed to allow for modulation, and the resulting data optically multiplexed before the subsequent transmission through fibres with EDFA amplification. At the receiver, each channel was demultiplexed before reception. ECL, edge-coupled laser, WSS wavelength-selective switch. Rx receiver.

## II. EXPERIMENT

*a)*        *CMOS compatible micro-comb source*

The micro-ring resonator (MRR) for comb generation was fabricated using CMOS compatible processes [33-50] with doped silica glass waveguides, which features low linear loss (~0.06 dB.cm−1), a moderate nonlinear parameter (~233 W−1.km−1), and negligible nonlinear loss that leads to an ultra-high nonlinear figure of merit. The MRR had a cross-section of 3×2 μm and radius ~592 μm, yielding an FSR of 48.9 GHz and a Q factor >1.5 million. The dispersion of the TM mode was designed to be anomalous in the C band with a jump at ~ 1552 nm brought about by the mode crossing. The bus waveguides of the MRR were directed to on-chip mode converters, then coupled to a single-mode fibre array, with a fibre-chip coupling loss of ~0.5 dB per facet. While statistical studies of fabrication yield are outside the scope of this work, we note that our platform is fully CMOS compatible using stepper mask aligners on full wafers [46]. Further, our low index contrast (core index = 1.7), results in larger waveguide dimensions which in turn makes them less sensitive to fabrication error. Our typical yields for FSR and Q factor are high – above 90%, and mode-crossings are not a challenge.

The soliton crystal device and the soliton crystal comb spectra are shown in Figures 2,3. The microcomb had a 48.9 GHz FSR, producing a soliton crystal output with a spectrum spanning across > 80 nm while pumping at 1.8 watts of CW power at a wavelength of 1550nm. The soliton crystal micro-comb was preceded first by the primary comb and displayed a variation in comb line powers at < +/- 0.9 dB, for ten different incidents of initiation, and was achieved by sweeping the wavelength manually from 1550.300 - 1550.527 nm. This clearly proves the micro-comb turn-key generation repeatability for our devices. Out of the total number of generated comb lines, eighty were chosen from the 3.95 THz, 32 nm wide C-band window at 1536 – 1567 nm.

The micro-comb used in the study was generated from the doped silica double-bus micro-ring resonator described above, packaged with a fibre array connection to all four device ports. We pumped the ring with a CW external cavity laser (Yenista Tunics – 100S-HP) at an output power of 15 mW, which was then amplified to 1.8W in a polarization maintaining erbium doped fibre amplifier (EDFA) (Pritel PMFA-37). Only the TM mode of the resonator oscillated in a soliton crystal state, hence the pump polarization was tuned to match this mode. As indicated in Figure 2a, we inserted the pump light into the 'through' port and collected light from the corresponding 'drop' port. The MRR chip was mounted on a Peltier cooler, monitored by a standard NTC temperature sensor. The temperature was maintained with a thermo-electric cooler (TCM-M207) at 25oC, within 0.1C of accuracy. The laser was set to standard running mode, with no extra steps made to stabilise the output frequency. Soliton crystal generation was achieved by automated wavelength tuning, in turn reducing the system complexity compared to other micro-comb generation schemes [24]. We measured the internal conversion efficiency of our soliton crystals to be 42% for the whole spectrum, and 38% when selecting the 80 lines over the C-band, highlighting that over 90% of our available comb power is compatible with standard C-band equipment.

While self-localised DKS waves require complex dynamic pumping schemes to initiate, soliton crystals are generated from a fundamentally different process, although both are described by the Lugiato-Lefever equation [2]. Soliton crystals are naturally formed in micro-cavities that display the appropriate form of mode crossings, without the need for complex dynamic tuning mechanisms. They were termed soliton crystals due to their crystal-like profile in the angular domain within the micro-ring resonators [11]. The formation of micro-combs is intimately related to the detuning between the pump wavelength and a resonance of the micro-photonic resonator [2]. To generate coherent and low noise micro-combs, the pump wavelength is swept from the blue to the red side of the resonance. This first excites primary combs, typically with line spacings of many resonator FSRs, followed by unstable chaotic combs with high intensity noise, before finally inducing solitons in the resonator. However, these soliton states typically have much lower intra-cavity power than the preceding chaotic states, therefore, as the soliton state is initiated, the resonances shift due to thermal effects and the soliton states become lost. Techniques such as fast wavelength sweeping [3, 24, 25] and power kicking can successfully capture soliton states, but this significantly increases the complexity and footprint of the system due to the need for external swept-frequency RF sources and modulators. Deterministic soliton generation cannot generally be achieved by pre-determined tuning into a resonance but requires instead auxiliary stabilization systems [54, 55]. The high performance of SCs including both robustness and efficiency [11, 52] is understood. It stems from the fact that in SCs the resonator is virtually completely filled with solitons making the intracavity energy very close to the chaotic state, whereas the energy of DKS states largely resides in the CW background rather than the single soliton pulse. Because soliton crystals are tightly packed systems of self-localized pulses, they have more than ten times higher intra-cavity power than the DKS regime — in fact, close to the power of the spatiotemporal chaotic states [11, 26]. Therefore, when sweeping the pump wavelength, switching between chaotic and soliton crystal states does not introduce significant changes in intra-cavity power. Hence, soliton crystals do not suffer from thermal detuning due to the characteristic 'soliton step' observed for single soliton states. This has the important consequence that they can be initiated through adiabatic pump wavelength sweeping - far simpler than what is required to initiate and maintain DKSs. As such, soliton crystals are highly robust and can provide stable micro-combs without the need for complex feedback systems [26, 52]. This intrinsic robustness is central for enabling the use of micro-combs outside of the laboratory. While soliton crystal micro-combs have been successfully exploited for microwave photonics [56], their potential for coherent optical communications has only been reported recently [26].

The evolution of a soliton crystal micro-comb while tuning the pump laser manually is shown in Figure 2. The open loop generation of the soliton crystal comb along with the low (~50 GHz) line spacing and wide bandwidth of the high conversion efficiency first 'lobe' (4 THz) makes these combs attractive for application to compact, high rate transceivers for optical communications. We note that it is also possible to generate either a different soliton crystal state (Figure 3a) or a chaotic state (Figure 3b), via slow, manual wavelength tuning. For comparison the actual soliton spectrum used in the experiments is shown again in (Figure 3c). The automatic wavelength sweeping over a pre-determined wavelength range (1550.300 – 1550.527 nm) to generate the wanted soliton crystal state was achieved by wavelength tuning the laser in 1pm steps, with half a second between each step. This tuning was achieved by simply setting the wavelength to different values on a Yenista Tunics T100HP, remotely controlling the unit via GPIB through Python. The tuning rate was found through trial and error to allow for reproducible generation of the desired soliton crystal state. While the generation of this state seems robust and repeatable, we make no firm claims to deterministic generation of soliton crystal states, as we have not modelled the generation of the desired state with thermal terms introduced into the LLE, which would allow for the simulation of the effects of thermo-optic chaos [2]. We note that there have been successful demonstrations of deterministic micro-comb generation, although these involve auxiliary systems which greatly increase overall system complexity [54, 55]. In order to estimate the internal conversion efficiency from the pump to comb line within the micro-ring resonator, we analyse the relative powers of the pump line and comb lines emitted from the drop port of the device. Measurement of light from the drop port of our 4-port, dual-bus device accurately reflects the light within the resonator. We define internal conversion efficiency as the power ratio between the pump line at 1551.05 nm and the comb lines, consistent with [1-4]. Analysing the spectrum gathered on a standard OSA running at 0.06 nm resolution, we measure an internal conversion efficiency of 42.1%. This compares favourably with dark [10] (20%) and bright [3, 4] solitons (< 0.6%), as expected. Clearly, when taking other factors into account, such as coupling loss, proportion of comb lines used, etc., this will naturally drop, as it will for all devices. We do note that our chips are fibre pigtailed with on-chip mode converters, resulting in extremely low fibre-chip coupling loss < 0.5dB. Further, after selecting the 80 comb lines over the C-band our internal efficiency was 38.1%. This highlights that our soliton crystal state provides the majority of its power (> 90%) in a useful bandwidth for standard C-band optical communications equipment.

Once established, the robustness of soliton crystal states to relatively large laser frequency drifts was verified by observing that the same stable oscillation state was maintained, and with a maximum power variation of only +2.4 dB to -1.7 dB for the 80 lines used for the transmission experiments even while tuning the pump laser by more than 12 pm (1.5 GHz) – much larger than the actual variation of the pump laser wavelength in the transmission experiments. We measured the power stability of the comb lines (Figure 4) over 66 hours, with spectra captured every 15 minutes with only open loop control (standard thermo-electric controllers) and without manual intervention to stabilise the micro-ring resonator chip. Figure 4a shows the measured spectrum for the 80 C-band comb lines along with the standard deviation (SD in dBm - given by the error bars), showing a relative SD (in dB, Figure 4b) of about -14dB over the 66 hour period. Conceptually, these measurements demonstrate that the fiber-coupled MRR device can operate as a separate, independent, plug-in element that multiplies the number of coherent carriers produced by an independent laser by 80 times. We emphasize, however, that in fact it is the net superchannel transmission that is the ultimate test of the comb's performance – not the power stability measurements – and so OSNR is far more important. As long as the required OSNR is maintained, high fidelity data transmission will be supported, which was the case in our experiments. Further, due to the self-stabilizing nature of soliton states in micro-resonators [2, 3], there are fundamental reasons to expect even greater open-loop long-term stability than we have shown in our current measurements. Finally, any improvements to the system such as the addition of feedback control loops, the use of a pump laser with greater frequency stability, better design of the micro-ring resonator thermal stabilization, or the avoidance of comb flattening, would enhance our superchannel transmission performance even further.

In our experiments we used an external laser source and amplifier to pump the micro-ring used to generate the optical frequency comb. We note that recent demonstrations of hybrid integration of pump lasers to generate single DKS microcombs [55-59], as well as advanced techniques such as injection locking [57], can equally be applied to soliton crystals – in fact probably even more easily given their much simpler generation process.Hybrid integrated pump sources also yield much greater energy efficiency - the states in [55] were produced with 4 dBm (2.5 mW) pump power. Moreover, the external cavity structure shown in [55] is intrinsically compatible with the much less complex and slower tuning required by soliton crystals.

Integrated laser arrays can now produce high quality optical carriers. Although a full analysis comparing the different approaches is beyond the scope of this work, we make a number of comments. To support tightly spaced superchannel multiplexing with 80 separately integrated lasers with the same performance as micro-combs, it would require precise wavelength locking of all 80 lasers to compensate for variations due to fabrication error. Moreover, the footprint of 80 discrete high-quality lasers may be prohibitively large. Micro-combs, on the other hand, have been realized with hybrid integrated pump sources and have achieved ultralow thresholds <10 mW [57]. Microcombs are intrinsically locked to the micro-resonator free-spectral range, without any feedback requirements. They also have a small footprint, being based on a single high quality resonator and single pump laser source.

### b)   *Systems experiment*

We performed 2 experiments: the first across 75 km of single mode optical fiber in the lab and the second in a field trial using a metropolitan network in the greater Melbourne area, also based on standard SMF, which linked Monash University's Clayton campus to the RMIT campus in the Melbourne CBD. A map of the metropolitan network used for the system field trial is given in Figure 1, while the soliton crystal device and the comb spectra are shown in Figure 2. The experimental setup for the demonstration of high capacity optical data transmission is shown in Figure 5 (simplified overview) and in Figure 6 in more detail.

The transmission link was comprised of two fibre cables connecting labs at RMIT University (Swanston St., Melbourne CBD) and Monash University (Wellington Rd, Clayton). These cables were routed from the labs access panels, to an interconnection point with the AARNet's fibre network. The fibre links were a mix of OS1 and OS2 standard cables and include both subterranean and aerial paths. There was no active equipment on these lines, providing a direct dark fibre connection between the two labs. The total loss for these cables was 13.5 dB for the RMIT-Monash link and 14.8 dB for the Monash-RMIT paths. The cable lengths as measured by OTDR were both 38.3 km (totalling 76.6 km in loop-back configuration). At Monash, an EDFA was remotely monitored and controlled using a 1310 nm fibre-ethernet connection running alongside the C-band test channels. The comb was amplified to 19 dBm before launch, at Monash, and on return to RMIT. The installed network fiber for the field trial presented a different testing platform to the spooled fibres used in lab. Splices and connections along the link between the two labs provide a source of uncontrolled back-reflections and limit the amount of power that can safely be sent over the network given the risk of connector burns and even fibre fuses from reflective interfaces. Coupled with the higher losses of installed (legacy) fiber links, this provides a challenging platform for high spectral efficiency optical communications where maximising signal to noise ratio is key to enabling high capacities. Moreover, the operation over legacy fibre links covering typical suburban distances demonstrates that it is possible to leverage installed fiber infrastructure for next generation metropolitan/regional area systems, which have been experiencing a higher growth in required capacity than long-haul networks, actually surpassing them in 2017 [60 SM 3]. This is particularly important due to the cost of laying new fiber in installed ducting being on the order of $30k / mile [61 SM 1]. It also demonstrates the feasibility of system upgrades using micro-comb-based transceivers to extend the useful lifetime of installed fiber systems.

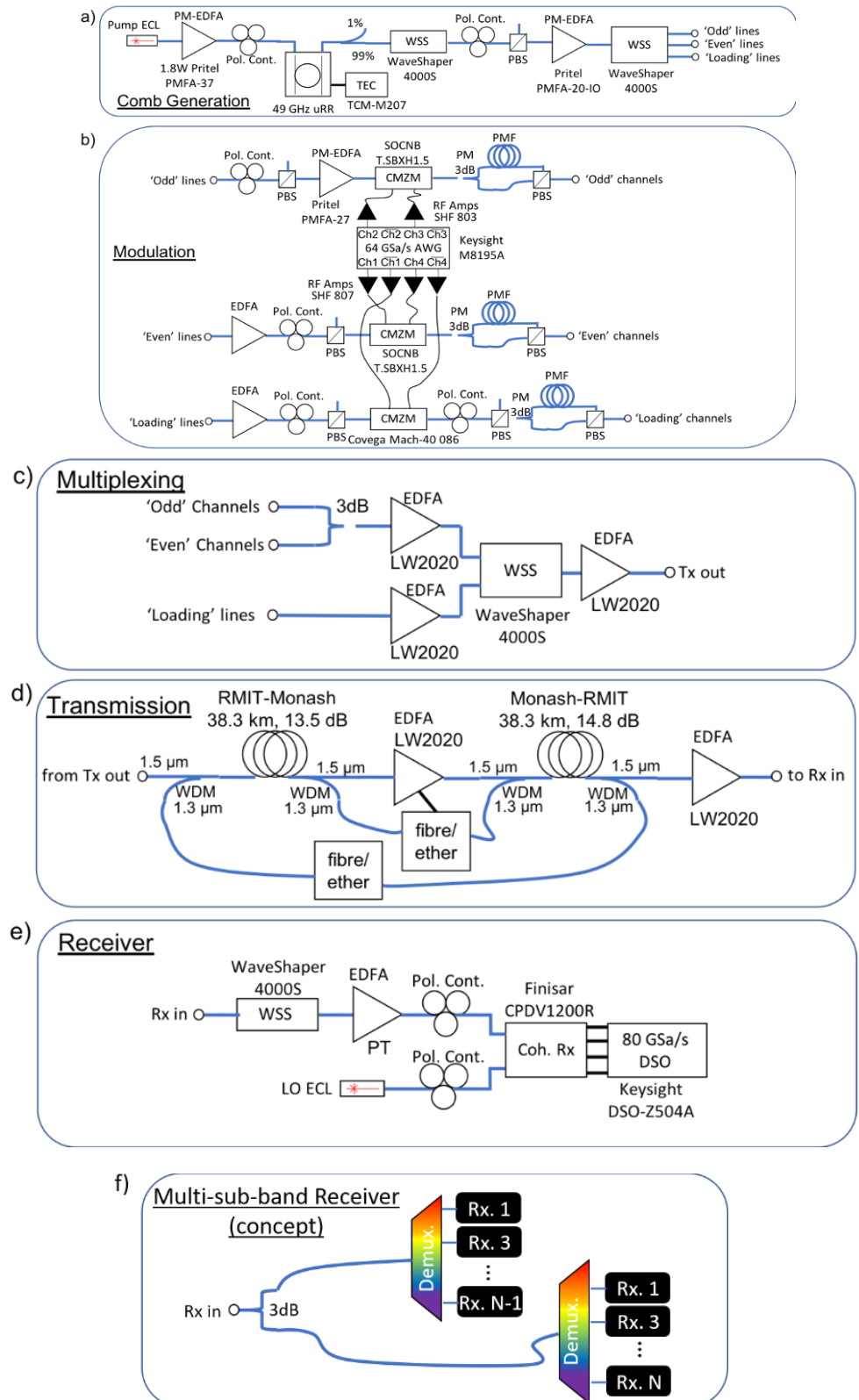

**Figure 6.** Experimental set-up. ECL: External cavity laser, PM-EDFA: Polarization maintaining EDFA, Pol. Cont.: Polarization controller, uRR: micro-ring resonator, TEC: Thermo-electric controller, WSS: Wavelength selective switch, PM 3dB: Polarization maintaining 50%/50% optical power coupler/splitter, PBS: Polarization beam splitter, PMF: Polarization maintaining fibre, CMZM: Complex Mach-Zehnder modulator (dual nested type), AWG: Arbitrary waveform generator, RF Amps.: microwave amplifiers, WDM: Wavelength division multiplexer (1310/1550 nm), fibre/ether: Gigabit Ethernet to fibre channel media converter, Coh. Rx: integrated optical hybrid and balanced photodiode coherent receiver, DSO: Digital storage oscilloscope (real-time). For the EDFAs, LW2020 denotes Lightwaves2020 MOABF17 optical amplifiers, PT is a Photonic Technologies 'Powerflat' amplifier. f) depicts the concept of a receiver for every sub-band.

The microcomb featured a 48.9 GHz FSR, producing a soliton crystal output with a spectrum spanning across > 80 nm while pumping at 1.8 watts of CW power at a wavelength of 1550nm. The soliton crystal micro-comb was preceded first by the primary comb and displayed a variation in comb line powers of < +/- 0.9 dB, for ten different incidents of initiation, and was achieved by sweeping the wavelength manually from 1550.300 - 1550.527 nm. This demonstrates the micro-comb turn-key generation repeatability.

Since the nonuniform spectrum of soliton crystal combs has been viewed as a weakness, we chose to flatten the optical frequency comb first so that all lines were of equal power, even though this is not necessary and actually introduces unnecessary impairments – both in our experiments and other micro-comb demonstrations (e.g. [24-26]). All comb lines were wavelength demultiplexed into separate waveguides and sent to separate modulators. It is then straightforward to adjust the comb line power by variable attenuators, amplifiers or even by varying the RF drive amplitude to the modulators. We implemented comb flattening for several reasons: i) to prove our system operation under the most demanding conditions, ii) to pre-empt the criticism that the nonuniform spectrum of SCs is a limitation, and iii) to facilitate easy comparisons with prior art. Since avoiding flattening would reduce impairments and improve our performance (by increasing the OSNR of the higher power comb lines and their ability to carry higher spectral efficiency modulation formats, and by eliminating the loss of the extra WaveShaper), it does not represent a limitation to SC based transmission.

The soliton crystal micro-comb was flattened in two stages by two independent programmable optical filters (Finisar WaveShaper 4000S). The WaveShapers had an insertion loss of 5dB each, in addition to any variable attenuation. The first had a static filter shape set to equalize each comb line to within about 1 dB of each other, to coarsely match the generic shape of the soliton crystal state we chose to use. The second programmable filter was set each time that a new soliton crystal state was initiated, to equalize the comb line powers to within < 1 dB of each other, although we note that it was often unnecessary to change the filter profile when generating a new soliton crystal. Spectral shaping in a WDM transceiver using a comb source involved minimal extra complexity since only the addition of attenuators after the WDM demultiplexer were required to route each comb line to a separate modulator. The comb was then amplified by a further polarization maintaining EDFA (Pritel PMFA-20-IO), before being divided for modulation. Prior to modulation, the optical signal-to-noise ratio (OSNR) of the individual comb lines was > 28 dB. Comb flattening is not necessary either in our experiments or other micro-comb demonstrations [24-26], since all comb lines are typically wavelength demultiplexed into separate waveguides and sent to separate modulators. It is then straightforward to adjust the comb line power by variable attenuators, amplifiers or even by varying the RF drive amplitude to the modulators. In fact, we expect better performance without comb flattening since the higher power comb lines would need less attenuation and/or amplification before modulation, resulting in a higher OSNR, while the lower power comb lines would have essentially the same performance as reported here. Furthermore, using the raw spectrum would avoid the loss of the extra Waveshaper. Therefore, avoiding flattening (working with the raw spectrum) would in fact yield even higher system performance.

Out of the total number of generated comb lines, eighty were chosen from the 3.95 THz, 32 nm wide C-band window at 1536 – 1567 nm. The spectrum was then flattened using a WaveShaper. Following this the number of wavelengths was doubled to 160, corresponding to a 24.5 GHz spacing, to increase the spectral efficiency. This was accomplished with a single sideband modulation technique that generated both even and odd channels that were not correlated. We then grouped six wavelengths, with the rest of the bands supporting data loaded channels based on the same even-odd structure. We were able to use a record high order 64 QAM coherent modulation format that modulated the whole comb at a baud rate of 23 Giga-baud, that achieved 94% utilization of the available spectrum. After the spectrum was flattened with the WaveShaper, the number of wavelengths doubled to 160, corresponding to a 24.5 GHz spacing, to increase the spectral efficiency. This was accomplished with a single sideband modulation technique that generated both even and odd channels that were not correlated. We then grouped 6 wavelengths, with the rest of the bands supporting data loaded channels based on the same even-odd structure. We used 64 QAM coherent modulation at 23 Giga-baud, with 94% utilization of the available spectrum.

*c)*    *Transmitter*

The experimental setup is shown in Figures 5 and 6. The transmitter used 3 separate complex Mach-Zehnder modulators to provide both odd and even test bands, as well as a loading band. The comb lines for each of these bands were split using another programmable filter (Finisar WaveShaper 4000S) and were then amplified before modulation. Three tones separated by 98 GHz around the selected test frequency were directed to two separate modulators (Sumitomo Osaka Electric Company New Business T.SBXH1.5-20 ). The two modulators were driven at a symbol rate of 23 Gbd, providing a per sub-band line rate (i.e. excluding overheads) of 23 Giga-symbols/s x 6 bits/symbol x 2 polarizations = 276 Gb/s. The sub-bands were shifted by 12 GHz from the optical carrier, with one modulator providing a sideband down-shifted from the optical carrier, and the other an up-shifted sideband. This enabled higher fidelity modulation than simple 46 Gbd single-carrier modulation, given the transceiver noise limitations we had in our system. The odd and even bands were decorrelated by adding a delay with an extra length of optical fibre in the 'even' path. A third modulator (Covega Mach-40 806) was used to modulate the loading bands which consisted of a two Nyquist sub-carrier modulation scheme to mimic the structure of the odd and even test bands. The two bands were driven by pairs of the positive and negative differential outputs of the AWG (Keysight M8195A, 65 GSa/s, 25 GHz bandwidth), while the loading channels were driven by a separate independent output pair. The modulating waveforms were set to provide 64 QAM signals, pulse shaped by a 2.5% roll-off RRC filter, running at 23 Gigabaud. On a 49 GHz grid, this provided a 94% spectral occupancy. The modulator optical outputs were each passed through a polarization maintaining 3 dB power splitter, one output being delayed by a few meters of optical fibre and then rotated by 90o using a polarization beam splitter/combiner. This provided emulation of polarization multiplexing by delay de-correlation. The odd, even and loading bands were all de-correlated from each other by means of different fibre delays of a few meters. The odd and even channels were passively combined with a 3-dB power splitter, to maintain the pulse shape of the central channels. The combined test and loading bands were multiplexed by a further programmable filter (Finisar WaveShaper 4000S). The roll-off of the filters from this device did affect the outer channels of the test band and the neighbouring channels in the loading channels. After multiplexing the fully modulated comb was amplified to a set launch power. The Tx DSP is described below.

The transmission section also shows the control channel used to monitor and control the EDFA hosted remotely in the Monash labs. This is provided by an Ethernet over fibre link, using a FS.com UM-GADF40 media converter at either end of the link. Control and test channel separation is provided by 1310/1550 nm WDM. The comb line OSNR is an important factor in determining the performance of optical frequency combs. We measured the in-band OSNR, accurately using a 150 MHz resolution optical spectrum analyser (Finisar WaveAnalyzer), at several points. We calibrated the WaveAnalyzer against a standard OSA working with 0.1 nm resolution through noise loading a single laser line. Directly after the microcomb, the OSNR was > 33 dB; After the amplifier between the WaveShaper shaping stages (Pritel PM-20-IO), the OSNR was 30 dB, and directly before the modulators on the odd and even 'test-band' arms, it was 28 dB. Note that both after the comb amplifier and directly before the modulators, the optical noise was co-polarized with the comb line. The WaveShapers had an insertion loss of 5dB each in addition to any variable attenuation.

Offline signal generation and reception is performed in MATLAB. The transmitter side defined a frame 218 samples long, to match the memory depth of the Keysight M8195A AWG. The frame consisted of a short sequence of zeros, used for system diagnostics (visually) from the oscilloscope waveforms, followed by a 400 symbol BPSK sequence which was used to denote the start of the data packet. Next, the rest of the samples occupied by a waveform were generated from random integers ranging from 0 to M-1 (where M is the QAM level modulated). The data packet consisted of 64-QAM symbols, where the random data was mapped to the constellation points using Gray coding. The frame was generated at 1 Sa/symbol, then up-sampled with zero padding between symbols to 2 Sa/symbol. The up-sampled waveform was filtered by a root-raised cosine (RRC) filter, set for a roll-off of 2.5% (beta = 0.025), with out-of-band attenuation set to 25 dB (filter defined by 'fdesign.pulseshaping'). This filtered signal was then up-shifted in frequency by 12 GHz. Finally, the shifted signal was resampled to the AWG sample rate. This was done using the 'resample' function that implemented an upsample-FIR-downsample script. The real and imaginary parts of the resulting waveform were sent to the AWG.

When modulating the odd-frequency bands (shifted by +12 GHz from the optical carrier) were generated by adjusting the bias on the CMZM to provide a +90-degree phase shift between the two nested MZMs in that device. For the even channels (shifted by -12 GHz from the optical carrier), this bias was set to -90 degrees. This enabled the normal and inverted outputs of the AWG to be used to generate the two sidebands, which were subsequently delay decorrelated through a length of fibre in the even arm. Independent waveforms were generated for the loading channels, which were modulated with a dual-sideband signal such that the generated loading channels emulated the combination of the odd and even test channels. Given that the electro-optic bandwidth of the Covega Mach-40 modulator was limited (much lower than for SOCNB modulators), we used a pre-emphasis filter to modify the driving waveforms so that a flat spectrum was generated for the loading channels. The pre-emphasis filter was generated as an amplitude-only filter with the filter shape derived from measuring the generated signal spectrum with a 150-MHz resolution optical spectrum analyser (Finisar WaveAnalyzer), and then inverting the measured spectral shape as a pre-emphasis filter. No pre-emphasis was used for the test channels.

### d)    *Receiver stage*

The receiver stage architecture is shown in Figure 6. Before photo-detection, the signal was filtered by a programmable optical filter (Finisar WaveShaper 4000S) set to a 35 GHz passband, in order to select the channel to be measured. The 35 GHz passband was found to be an optimal setting in experiment (see below for more detail). Note that it was not possible for this device to provide a filtering response that matched the channel shape, as these devices typically have a 10 GHz optical transfer function, limiting filter roll-off [63]. Moreover, since we used a 2.5% RRC shaping filter in the transmitter side DSP, if the set receiver side optical filter did not exhibit a flat passband over the signal bandwidth, the signal would have been degraded. Additionally, using too broad a filter would waste receiver dynamic range by partly detecting neighbouring sub-bands [64]. As such, we swept the receiver side filter passband setting, and found that 35 GHz was an optimum value.

The output of the filter was amplified to approximately 10 dBm before being directed into a dual-polarization coherent receiver (Finisar CPDV1200, 43 GHz bandwidth). A local oscillator was provided by an Agilent N7714A laser tuned close to the comb line of interest, at 16 dBm of output power. The photo-detected signals were digitized by the 80-giga-samples-per second (GSa/s), 33-GHz bandwidth inputs of a Keysight oscilloscope (DSO-Z504A, 50 GHz, 160 GSa/s). The digitized waveforms were forwarded to a PC for offline digital signal processing. The digital signal processing flow started with renormalization, followed by overlap-add chromatic dispersion compensation, then a spectral peak search for frequency offset compensation, followed by frame synchronization using a short BPSK header, before final equalization. As the specific fibre types used along the link are not well known, the level of chromatic dispersion compensation was estimated through analysing the header correlation peak height. Equalization occurred in two stages, with a training-aided least-means-squared (LMS) equalizer performing pre-convergence, the taps of which were sent to a blind multi-modulus equalizer. After equalization, a maximum-likelihood phase estimator was used to mitigate phase noise, before the signal was analysed in terms of BER, EVM and GMI. The signal was recovered at the receiver with a standard offline digital signal processor (DSP).

The receiver side DSP removed the mean component of the waveforms, running via a Gram-Schmidt orthogonalization method to compensate for receiver I/Q imbalances, followed by overlap/add dispersion compensation with blocks of 1024 samples, and resampled down to 2 Sa/symbol using MATLABs 'resample' function. A spectral peak search was performed to determine the frequency offset by finding the minimal residual carrier left after modulation. The waveform was then filtered by concatenating a static RRC filter to the pulseshaping filter, matching the one used in the transmitter. Frame synchronization was performed by finding the correlation peak between the received waveform and the sent synchronization header. Equalization was performed using an 81-tap filter. For convenience, we used the LMS algorithm to initialize a multi-modulus algorithm, as the training-based LMS stage performed source separation without the potential for single polarization convergence. Such a convergence can artificially improve the performance of systems like ours that use delay decorrelation-based polarization multiplexing emulation. We used 10,000 data symbols for training, representing approximately 9% of the total data packet. Note that we were limited in packet size to 4.1 μs (94208 symbols) by the memory depth of our AWG. We expect the channel to remain stable over a longer period of time, such that the training overhead used here would in practice become negligible (for example, a 41 us frame would reduce training overhead to <1%). Moreover, we also post-processed our data using a blind CMA algorithm as a pre-equalization stage, and this resulted in negligible overall performance reduction (less than about 2.5%). Some captures did not converge as well, which we attribute to our particular equalizer implementation and not a fundamental shortcoming of the blind algorithms. We believe that careful optimization of the blind equalization stage would result in negligible penalty compared to the data-aided equalizer. The multi-modulus algorithm worked on a blind radial decision basis, and the filter arose from this algorithm used to equalize the received waveforms. As the DSO and AWG clocks were independent and did not share a common reference, the equalizer was also used as a stand-in for clock recovery. After equalization, a maximum-likelihood based phase estimator was used to provide a phase noise correction, running over a 16-bit averaging window. After phase correction, BER, Q2 and GMI were calculated. Q2 was inferred from error-vector magnitude (EVM), as 20.log10(sqrt(1/EVM)). EVM was calculated from the difference between the magnitude of the sent and received signals. GMI was calculated using the calcGMI.m script provided at https://www.fehenberger.de/#sourcecode.

We also provide additional data on the sensitivity of the 23 Gbd, 64 QAM signal to local oscillator OSNR to illustrate the potential of soliton crystal states to perform as a multiwavelength local oscillator. Figure 9 shows the results for the signal quality factor and generalized mutual information (GMI) versus local oscillator OSNR. To make a more direct comparison with our comb, the local oscillator was noise loaded. The OSNR was then measured and filtered by a wavelength selective switch with a bandpass setting of 10 GHz, amplified, and then passed through a polarizing beam splitter before being launched onto the coherent receiver as a local oscillator. We find that there is minimal penalty for the received signal at a comb line OSNR >27 dB, observed in the experiment. We measure a 1 dB penalty in Q2, or a reduction in GMI of 0.3 bits/symbol, which corresponds to an effective rate reduction of about 2.5%.

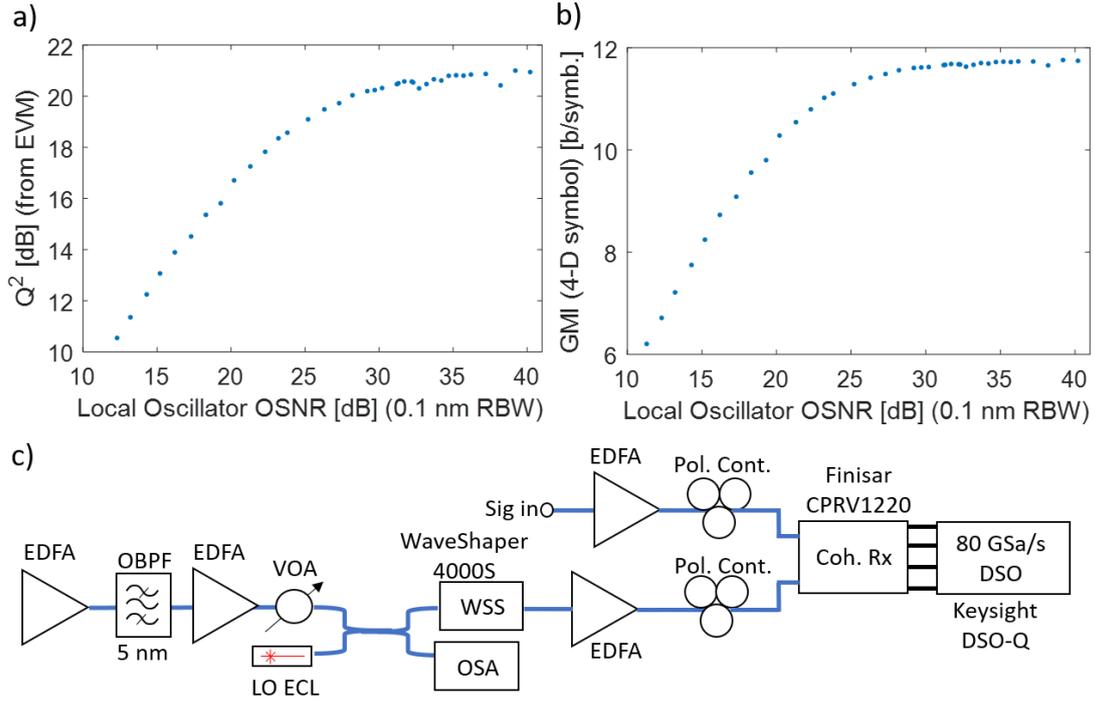

**Figure 7.** Signal tolerance to local oscillator (LO) optical signal to noise ratio (OSNR). a) Singla quality factor ($Q^2$, as extracted from error vector magnitude as $20\log_{10}[1/EVM^2]$) and b) generalized mutual information per 4-D symbol (GMI) against local oscillator quality factor. OSNR is loaded onto the local oscillator, then filtered by a WSS set to provide a 10 GHz passband, as shown as c). This filtering set-up mirrors that used for the frequency comb. OBPF: Optical band pass filter, WSS: Wavelength Selective Switch, EDFA: Erbium-doped fibre amplifier, ECL: External cavity laser, OSA: Optical spectrum analyser, DSO: Digital Sampling Oscilloscope, Coh. Rx: Coherent receiver.

### III. RESULTS AND DISCUSSION

Figures 8, 9 show the results of the experiments. As discussed, we conducted two transmission experiments, sending data over 75 km of single mode fibre in the laboratory as well as in a field trial across an installed metropolitan-area single-mode fibre network connecting the Melbourne City campus of RMIT and the Clayton campus of Monash University, spanning the greater metropolitan area of Melbourne. From the generated micro-comb, 80 lines were selected over the telecommunications C-band (32 nm wide, 3.95 THz window from 1536 – 1567 nm) which were then flattened with a WaveShaper and the number of wavelengths doubled to 160 (equivalent to a spacing of 24.5 GHz) to optimize the spectral efficiency with a single-sideband modulation scheme to generate odd/even decorrelated test channels. We then combined a test band of six channels, with the remaining bands providing loading channels having the same odd-and-even channel structure. We used a record high order format of 64 QAM to modulate the entire comb at 23 Giga-baud, resulting in the utilization of 94% of the available spectrum.

Spectra of the comb at key points are given in Figure 8 a-c. Figure 8d shows constellation diagrams for the signal at 194.34 THz. In back-to-back configuration (i.e. with transmitter directly connected to receiver) we measured signal quality (Q2, from error vector magnitude) at almost 18.5 dB, dropping to near 17.5 dB when transmitting the fully modulated comb through the test links.

Figure 9a shows the transmission performance using the bit error ratio (BER) for each channel as a metric, showing the 20% threshold for soft-decision forward error correction (SD-FEC), a common benchmark for performance, using a proven code, at a BER of $4\times10^{-2}$ [65]. Three scenarios were investigated: i) a direct connection between the transmitter stage to the receiver (back-to-back, B2B) and after transmission through ii) in-lab fibre and iii) over the field trial network. Transmission globally degraded the performance of all channels, as expected.

All results were below the given FEC limit, but since using SD-FEC thresholds based on BER is less accurate for higher order modulation formats and for high BERs [66], we additionally used generalized mutual information (GMI) to calculate the system performance. Figure 9b plots the GMI for each channel and its associated SE, with lines given to indicate projected overheads. We achieved a raw bitrate (line-rate) of 44.2 Tb/s, which translates to an achievable coded rate of 40.1 Tb/s (in B2B), dropping to 39.2 Tb/s and 39.0 Tb/s for the lab and field trial transmission experiments, respectively. These yielded spectral efficiencies of 10.4, 10.2 and 10.1 b/s/Hz.

This data rate represents an increase of nearly 50% (see below) over the highest reported result from a single integrated device [24]. Even more importantly the SE is enhanced even more, being a factor of 3.7. This is notable considering that we performed our experiments under the most demanding conditions, including not using any closed-loop feedback, stabilization or elaborate initiation schemes, as well as with the use of full comb flattening (equalization). Even though it is not necessary, we implemented flattening mainly to address possible concerns that the nonuniform spectra of soliton crystals might pose a limitation. Given that we achieved our results with flattening, and avoiding it would eliminate the impairments it introduced, thus improving our performance, it therefore does not represent a limitation. The same argument holds true for closed-loop feedback control of the micro-comb.

The record high spectral efficiency and absolute bandwidth that we achieve were enabled by the very high conversion efficiency we achieved between the pump and the soliton crystal comb lines [11, 52]. Again, as mentioned this results from the very small power step in the cavity that occurs when the soliton crystals are generated from the chaos states.

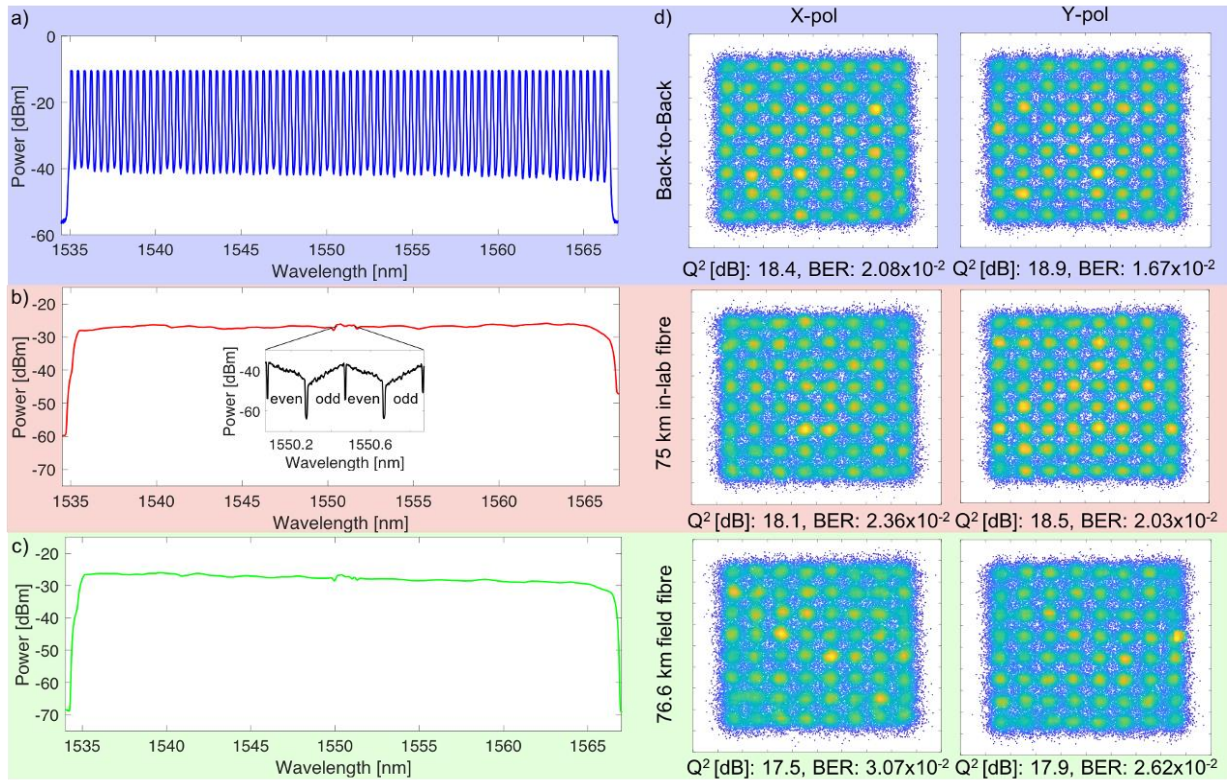

**Figure 8.** Spectra of the soliton crystal frequency comb after flattening (a), modulation and transmission through either 75 km spooled in-lab fibre (b) or through the field-trial link (c). The spectrum (a) is measured with 12.5 GHz resolution to resolve the individual comb lines, while (b) and (c) are plotted at 50 GHz resolution to illustrate average channel powers. Flattening equalised the comb line power to within 1 dB. After modulation and amplification, the channels were shaped by the EDFA gain spectrum. The inset in (b) depicts the test channel spectra captured with a 150 MHz resolution optical spectrum analyser, highlighting the odd and even sub-bands modulated onto each comb line in the test band. d) Constellation diagrams for a comb line at 193.4 THz (1550.1 nm) for both X- and Y-polarization channels. 'Back to back' denotes the transmitter directly connected to the receiver, '75 km in-lab fibre' indicates reception after transmission through 75 km of spooled fibre inside the lab, while '76.6 km field fibre' denotes reception after transmission through the field-trial link. BER and $Q^2$ related to the constellations are noted.

We only used the telecom C-band, and yet the bandwidth of the microcomb was larger than 80 nm. Therefore, wavelengths in both the L (1565-1605 nm) and even S (1500-1535 nm) bands could easily be used. In fact, even broader bandwidths can be achieved by increasing the power, by varying the wavelength of the pump, by engineering the dispersion or further methods. This would yield an increase of more than a factor of 3 in total bandwidth, resulting in >120 Terabits per second using only a single source.

Achieving even lower spacings, or FSRs, with micro-combs would yield yet higher SEs since the quality of the signal increases for smaller baud rates. This may result in a smaller overall comb bandwidth however. For our experiments, the use of single sideband modulation allowed multiplexing two channels using one single wavelength, which cut the comb spacing by a factor of 2 while enhancing the back-to-back performance that was limited by transceiver noise. This was made possible by the stability of the soliton crystals. Conversely, electro-optic modulation has also been used to sub-divide the micro-comb repetition rate, and this would also create broader comb bandwidths. This, however, would require locking the comb FSR spacing to an external RF source, although this is feasible since sub megahertz stabilization of microcombs has been achieved [67-69]. Furthermore, increasing the comb conversion efficiency by using a newly discovered class of soliton, called laser cavity-soliton micro-combs [34] will offer a powerful way to increase the system capacity as well as the quality of the signal even further. For recently installed networks, our approach can easily be complemented by using spatial division multiplexing based on multiple core fibre [70], yielding bandwidths of more than a petabit per second using a single microcomb. Our results join the many breakthroughs achieved with microcombs, and in particular using soliton crystal combs. These particularly include our applications of soliton crystals to RF and microwave signal processing [71-92]. This work presented here is the most challenging demonstration ever reported for micro-combs in terms of ease of generation, coherence, stability, noise, efficiency, and others, and is a direct result of the superior soliton crystal microcomb qualities.

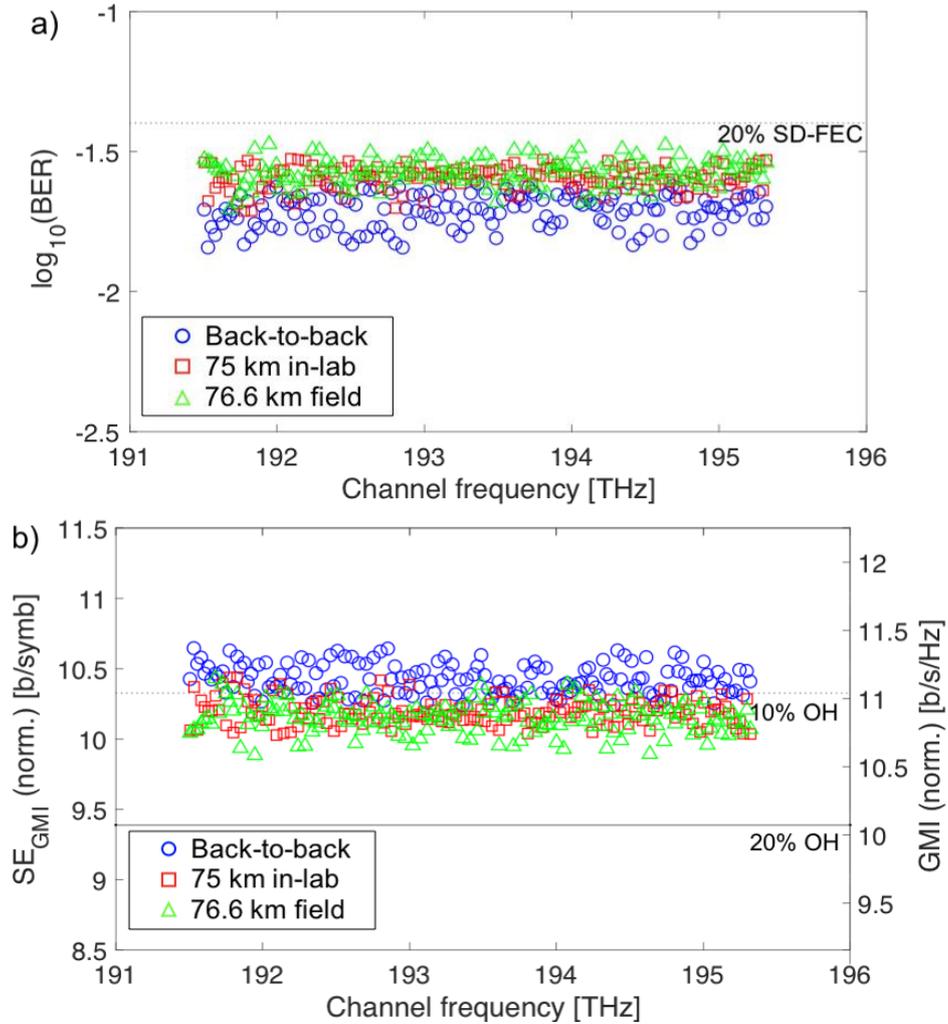

**Figure 9.** a) BER for each comb line. Blue circles points indicate performance of channels in B2B configuration, red squares dots are for performance after transmission through 75 km of in-lab spooled fibre, while green triangles are after transmission through the 76.6 km installed metropolitan-area fibre link. An indicative FEC threshold is given at $4\times10^{-2}$, corresponding to a pre-FEC error rate for a 20% soft-decision FEC based on spatially-coupled LDPC codes [99] (dashed line). After transmission, all channels were considered to be error-free, b) GMI and spectral efficiency measured for each comb line. GMI was calculated after normalization to scale measured constellations in order to account for received signal-to-noise ratio (SNR). Lines are for 20% and 10% overheads. Spectral efficiency was derived from GMI, and the ratio of symbol rate to comb spacing. GMI indicates a higher overall capacity than BER with the indicated SD-FEC threshold, as GMI assumes the adoption of an ideal code for the system. For B2B, GMI (SE) varied between 11.3 b/symb. (10.6 b/s/Hz) and 10.9 b/symb. (10.3 b/s/Hz). After in-lab fibre transmission, the achievable per-channel GMI (SE) varied between 11.0 b/symb. (10.4 b/s/Hz) and 10.7 b/symb. (10.1 b/s/Hz), with the same range observed for the installed field-trial fibres. We estimate the overall capacity from the sum of the GMIs, multiplied by the symbol rate.

## IV. PERFORMANCE COMPARISON

Here, we provide performance comparisons against record capacity results in a range of fibres. Table 1 summarizes key results from the literature comparing the various system performance metrics for demonstrations based on both integrated sources and rack-mounted equipment in both standard fibre or calculated on a per-mode basis for multicore fibre. See footnote [93] for details on how we calculate our performance metrics. Prior to our work, the best result (per core) was from [24], where, a single microcomb supported 30.1 Tb/s over the C and L bands, when using a standard tuneable laser coherent receiver. This is the benchmark result that we compare our results to in the main text, since it is not only the best published result using a single micro-comb, but it closely resembles our experiment (single micro-comb at transmitter, single tuneable laser at the receiver as a local oscillator). Note that our system uses only the C-band [24], while we improve the data rate because of the higher spectral efficiency. Reference [24] provides three different system set-ups. The first is closest to our own demonstration, and we use this in the main text for a direct comparison. In this case a single micro-comb is used at the transmitter, and a single laser employed as a local oscillator at the receiver. In the second demonstration in [24], two frequency interleaved combs are used in the transmitter, which, although resulting in a higher overall data rate, resulted in a lower per-comb spectral efficiency, and consequently a lower overall per-comb data rate than the single-comb demonstration. In the third demonstration in [24], a microcomb source is used as a local oscillator in the receiver, with a single comb employed at the transmitter side. In this case, overall system performance was improved to achieve a higher spectral efficiency and hence a higher overall rate, although lower than what we have demonstrated here.

High modulation formats have also been achieved with dark solitons [32], yet at a lower overall data rate, primarily due to the large comb line spacing that limits the spectral efficiency. In ref. [32], the same modulation cardinality (i.e. 64 QAM) was used; however, the high comb line spacing and relatively low symbol rate translated to a low spectral occupancy, resulting in a low spectral efficiency and aggregate data rate. The dark soliton state used in this demonstration also seems to require feedback to stabilize the state, increasing system complexity.

The work of [29] used a comb generator based on a benchtop pulsed seed laser source combined with waveguide spectral broadening. To provide a fully integrated system, this source would need to be on-chip. The focus in that experiment was using novel, proprietary multi-core fibre to achieve a 30-fold increase in bandwidth over the fibre in this spatially multiplexed system, to reach 0.66 Petabits/s. On a per-mode basis, Ref. [29] yields 25.6 Tb/s/mode, a lower per-mode capacity than this work and [24]. We note that both our approach and that of [24] are able to take advantage of SDM techniques to scale the overall bandwidth by using multi-core fibre. Although spectral utilization in Ref [29] was high, the achieved per-mode spectral efficiency is lower than our demonstration, and used 16-QAM modulation.

In contrast, soliton-based microcomb generation from an integrated hybrid chip has been demonstrated [57, 58]. For completeness, we also compare against record per-mode results for non-chip based sources - both benchtop-scale and rack mounted comb sources, and with traditional multiple discrete laser based WDM systems. The highest achieved per-mode data rate from a single comb used a non-integrated rack-mounted, or benchtop fibre-based comb [70] – ie., orders of magnitude larger (and more expensive) than chip-based microcombs, and which has not been realized in integrated form. There, the per-mode capacity across the C & L bands was 97.75 Tb/s, by using high cardinality modulation (64-QAM), high spectral utilization (24.5 Gbd on a 25 GHz grid) and efficient forward error correcting codes. This demonstration indicates that the expansion to the L-band with comb spacings similar to our own can enable even higher aggregate rates. The highest aggregate rate achieved over single mode fibre [94] used multiple discrete light sources. Again, that demonstration leveraged both C & L bands, although the achieved spectral efficiency is not significantly greater than in our demonstration, even when using 256-QAM relatively high code rate overheads were required (40-50%). This indicates that a single micro-comb source with quality similar to our own can provide performance comparable to multiple discrete laser sources. We note that the results presented in [94] were achieved using highly specialized fibres and with specialized amplification, neither of which are available in typical networks. The high aggregate rate they present was in part due to the full use of C- and L-bands, compared with the single C-band system we demonstrate here.

Our high spectral efficiency is partly enabled by the baud rate of the channels we modulate. The optimum baud rate in systems where OSNR is not the dominant performance limiter is the subject of on-going research. In systems with high OSNR, noise added by the transmitter and receiver limit performance, either from thermal noise from electrical components, quantization noise from the A/D and D/A converters, as well as noise and distortion from modulation and photodetection. This has proved important for superchannel reception [95, 96]. Also, we note that ultra-high spectral efficiency (>15 b/s/Hz) modulation has been based on low baud rates (between 6-10 Gbd) [97, 98]. It is on this basis that we suggest that combs with a lower FSR may enable a higher spectral efficiency, which should also improve the overall single-device data rate. The trade-off is that a reduction of line spacing results in a lower power per-line, impacting the individual comb line OSNR (if the same total bandwidth is preserved), which may ultimately pose a limit to the spectral efficiency as has been noted [67]. Understanding this trade-off is the subject of on-going research. The work presented here represents a new application for Kerr soliton crystal microcombs in addition to neural networks [71,72], microwave and RF signal generation and processing [73-92, 100-154] that will also benefit from the use of novel 2D materials [155-178].

**Table I**

| Line Rate | Net Rate | SE | Transmission | Reference |
|---|---|---|---|---|
| 30.1 Tb/s | 28.0 Tb/s | 2.8 b/s/Hz | 75 km SMF in-lab | Ref. [24], 1st |
| 27.5 Tb/s | 25.1 Tb/s | 2.6 b/s/Hz | 75 km SMF in-lab | Ref. [24], 2nd |
| 37.2 Tb/s | 34.6 Tb/s | 3.5 b/s/Hz* | 75 km SMF in-lab | Ref. [24], 3rd |
| 44.2 Tb/s | 40.1 Tb/s | 10.4 b/s/Hz | B2B (0 km) | This work |
| 44.2 Tb/s | 39.2 Tb/s | 10.2 b/s/Hz | 75 km SMF in-lab | This work |
| 44.2 Tb/s | 39.0 Tb/s | 10.1 b/s/Hz | 76.6 km SMF installed | This work |
| 4.8 Tb/s* | 4.4 Tb/s | 0.98 b/s/Hz* | 80 SMF km in-lab | Ref. [32] |
| 25.6 Tb/s* | 22.0 Tb/s* | 3.2 b/s/Hz* | 9.6 km, 30-core | Ref. [29] |
| 117.3 Tb/s* | 97.75 Tb/s | 9.8 b/s/Hz* | 31 km, 22-core | Ref. [70]** |
| 174 Tb/s* | 120 Tb/s | 10.8 b/s/Hz* | 70 km 'Z'-type SMF | Ref. [94]*** |

**Note** – results highlighted in yellow are not from chip-based sources but from rack mounted or benchtop based sources, included for completeness.

**Table 1.** Key systems performance metrics, per comb source used in the transmitter, and on a per mode basis. '*' indicates that this figure was not directly provided in the reference, and so is inferred from data provided. '**' indicates a demonstration using a commercial benchtop comb source, '***' indicates a traditional WDM result using multiple laser sources.

## V. CONCLUSIONS

We demonstrate a new world record for performance of ultra-high bandwidth optical transmission systems using a single optical source over standard optical fiber. We achieve this through the use of soliton crystal micro-combs that have a very low FSR spacing of 48.9GHz. Our achievement results from this record low comb spacing together with the efficient, broad bandwidth, and stable nature of soliton crystals, together with their CMOS compatible integration platform. Soliton crystal micro-combs are fundamentally low noise and coherent and can easily be initialised and operated using only very simple open-loop control that only requires commercially available components. Our results clearly show the ability of soliton crystal microcombs to achieve world record high bandwidths for optical data transmission over fibre in very demanding real-world applications.

8[44] A. Pasquazi, Y. Park, J. Azana, et al., "Efficient wavelength conversion and net parametric gain via Four Wave Mixing in a high index doped silica waveguide," Optics Express, vol. 18, no. 8, pp. 7634-7641, 2010.

[45] M. Peccianti, M. Ferrera, L. Razzari, et al., "Subpicosecond optical pulse compression via an integrated nonlinear chirper," Optics Express, vol. 18, no. 8, pp. 7625-7633, 2010.

[46] Little, B. E. et al., "Very high-order microring resonator filters for WDM applications", IEEE Photonics Technol. Lett. **16**, 2263–2265 (2004).

[47] M. Ferrera et al., "Low Power CW Parametric Mixing in a Low Dispersion High Index Doped Silica Glass Micro-Ring Resonator with Q-factor > 1 Million", Optics Express, vol.17, no. 16, pp. 14098–14103 (2009).

[48] M. Peccianti, et al., "Demonstration of an ultrafast nonlinear microcavity modelocked laser", Nature Communications, vol. 3, pp. 765, 2012. DOI:10.1038/ncomms1762

[49] A. Pasquazi, L. Caspani, M. Peccianti, et al., "Self-locked optical parametric oscillation in a CMOS compatible microring resonator: a route to robust optical frequency comb generation on a chip," Optics Express, vol. 21, no. 11, pp. 13333-13341, 2013.

[50] A. Pasquazi, M. Peccianti, B. E. Little, et al., "Stable, dual mode, high repetition rate mode-locked laser based on a microring resonator," Optics Express, vol. 20, no. 24, pp. 27355-27362, 2012.

[51] Lugiato, L. A., Prati, F. & Brambilla, M. Nonlinear Optical Systems, (Cambridge University Press, 2015).

[52] Wang, W., et al.., Robust soliton crystals in a thermally controlled microresonator, Opt. Lett., 43, 2002 (2018).

[53] Bao, C., et al., Direct soliton generation in microresonators, Opt. Lett, 42, 2519 (2017).

[54] Zhou, H., et al., Soliton bursts and deterministic dissipative Kerr soliton generation in auxiliary-assisted microcavities, *Light: Science and Applications*, **8**, 50 (2019).

[55] Kang, Z., et al., Deterministic generation of single soliton Kerr frequency comb in microresonators by a single shot pulsed trigger, *Opt. Express,* **26**, 326548 (2018)

[56] Xu, X., et al., Photonic microwave true time delays for phased array antennas using a 49 GHz FSR integrated micro-comb source, *Photonics Research*, **6**, B30-B36 (2018)

[57] Stern, B, et al., Battery-operated integrated frequency comb generator, *Nature,* **562,** 401-406 (2018)

[58] Raja, A.S., et al., Electrically pumped photonic integrated soliton microcomb, *Nat. Comms.,* **10**, 680 (2019)

[59] Pavlov, N.G., Narrow-linewidth lasing and soliton Kerr microcombs with ordinary laser diodes, *Nat. Photon.*, **12**, 694–698(2018)

[60] Cisco Corp., Cisco Visual Networking Index: Forecast and Trends, 2017–2022 White Paper, available at https://www.cisco.com/c/en/us/solutions/service-provider/visual-networking-index-vni/index.html#~complete-forecast .

[61] S. Grubb, et al., Field trials: Is it make or break for innovative technologies?, *Proc. OFC 2018*, S2A (2018)

[62] Shuto, Y., et al., Fiber fuse phenomenon in step-index single-mode optical fibers, *J. Quantum Electron.*, **40**, 1113-1121 (2004)

[63] C. Pulikkaseril et al., Spectral modelling of channel band shapes in wavelength selective switches, *Opt. Express*, **19**, 8458-8470 (2011)

[64] Y. Mori et al., Wavelength-division demultiplexing enhanced by silicon-photonic tunable filters in Ultra-Wideband Optical-Path Networks, *J. Lightwave Technol.*, DOI 10.1109/JLT.2019.2947709 (2019)

[65] K. Schuh, et al., Single Carrier 1.2 Tbit/s Transmission over 300 km with PM-64 QAM at 100 GBaud, Proc. Optical Fiber Communications (OFC), Th5B.5, San Diego, CA (2017).

[66] Alvarado, A., et al., Replacing the Soft-Decision FEC Limit Paradigm in the Design of Optical Communication Systems, J. Lightwave Technol. 34, 707 (2016).

[67] Torres-Company, V., et al.., Laser Frequency Combs for Coherent Optical Communications, J. Lightwave Technol. doi: 10.1109/JLT.2019.2894170 (2019).

[68] Obrzud, E., Lecomte, S. & Herr, T., Temporal solitons in microresonators driven by optical pulses, Nat. Photon., 11, 600 (2017).

[69] Papp, S.B., et al., Microresonator frequency comb optical clock, Optica 1, 10 (2014).

[70] Puttnam, B., et al., 2.15 Pb/s Transmission Using a 22 Core Homogeneous Single-Mode Multi-Core Fiber and Wideband Optical Comb, Proc. European Conference on Optical Communications (ECOC), PDP 3.1, Valencia (2015).

[71] X. Xu, *et al.*, "Photonic perceptron based on a Kerr microcomb for scalable high speed optical neural networks", *Laser and Photonics Reviews,* vol. 14, no. 8, 2000070, 2020. DOI:10.1002/lpor.202000070.

[72] X. Xu, *et al.*, "11 TOPs photonic convolutional accelerator for optical neural networks", Nature **589,** 44-51. 2021.

[73] J. Wu, X. Xu, T. G. Nguyen, S. T. Chu, B. E. Little, R. Morandotti, A. Mitchell, and D. J. Moss, "RF Photonics: An Optical Microcombs' Perspective," *IEEE J. Sel. Top. Quantum Electron.*, vol. 24, no. 4, pp. 6101020, Jul-Aug. 2018. DOI: 10.1109/JSTQE.2018.2805814.

[74] T. G. Nguyen *et al.*, "Integrated frequency comb source-based Hilbert transformer for wideband microwave photonic phase analysis," *Opt. Express,* vol. 23, no. 17, pp. 22087-22097, Aug. 2015.

[75] X. Xu, J. Wu, M. Shoeiby, T. G. Nguyen, S. T. Chu, B. E. Little, R. Morandotti, A. Mitchell, and D. J. Moss, "Reconfigurable broadband microwave photonic intensity differentiator based on an integrated optical frequency comb source," *APL Photonics*, vol. 2, no. 9, 096104, Sep. 2017.

[76] X. Xu, M. Tan, J. Wu, R. Morandotti, A. Mitchell, and D. J. Moss, "Microcomb-based photonic RF signal processing", *IEEE Photonics Technology Letters*, vol. 31 no. 23 1854-1857, 2019.

[77] X. Xu, *et al.*, "Broadband RF channelizer based on an integrated optical frequency Kerr comb source," *Journal of Lightwave Technology,* vol. 36, no. 19, pp. 4519-4526, 2018.

8
[102] Mengxi Tan, X. Xu, J. Wu, A. Boes, T. G. Nguyen, S. T. Chu, B. E. Little, R. Morandotti, A. Mitchell, and David J. Moss, "Advanced microwave signal generation and processing with soliton crystal microcombs", or "Photonic convolutional accelerator and neural network in the Tera-OPs regime based on Kerr microcombs", Paper No. **11689-38,** PW21O-OE201-67, Integrated Optics: Devices, Materials, and Technologies XXV, SPIE Photonics West, San Francisco CA March 6-11 (2021). DOI: 10.1117/12.2584017

[103] Mengxi Tan, Bill Corcoran, Xingyuan Xu, Andrew Boes, Jiayang Wu, Thach Nguyen, Sai T. Chu, Brent E. Little, Roberto Morandotti, Arnan Mitchell, and David J. Moss, "Optical data transmission at 40 Terabits/s with a Kerr soliton crystal microcomb", Paper No.11713-8, PW21O-OE803-23, Next-Generation Optical Communication: Components, Sub-Systems, and Systems X, SPIE Photonics West, San Francisco CA March 6-11 (2021). DOI:10.1117/12.2584014

[104] Mengxi Tan, X. Xu, J. Wu, A. Boes, T. G. Nguyen, S. T. Chu, B. E. Little, R. Morandotti, A. Mitchell, and David J. Moss, "RF and microwave photonic, fractional differentiation, integration, and Hilbert transforms based on Kerr micro-combs", Paper No. 11713-16, PW21O-OE803-24, Next-Generation Optical Communication: Components, Sub-Systems, and Systems X, SPIE Photonics West, San Francisco CA March 6-11 (2021). DOI:10.1117/12.2584018

[105] Mengxi Tan, X. Xu, J. Wu, A. Boes, T. G. Nguyen, S. T. Chu, B. E. Little, R. Morandotti, A. Mitchell, and David J. Moss, "Broadband photonic RF channelizer with 90 channels based on a soliton crystal microcomb", or "Photonic microwave and RF channelizers based on Kerr micro-combs", Paper No. 11685-22, PW21O-OE106-49, Terahertz, RF, Millimeter, and Submillimeter-Wave Technology and Applications XIV, SPIE Photonics West, San Francisco CA March 6-11 (2021). DOI:10.1117/12.2584015

[106] X. Xu, M. Tan, J. Wu, S. T. Chu, B. E. Little, R. Morandotti, A. Mitchell, B. Corcoran, D. Hicks, and D. J. Moss, "Photonic perceptron based on a Kerr microcomb for scalable high speed optical neural networks", IEEE Topical Meeting on Microwave Photonics (MPW), pp. 220-224,.Matsue, Japan, November 24-26, 2020. Electronic ISBN:978-4-88552-331-1.

**DOI:** 10.23919/MWP48676.2020.9314409

[107] Mengxi Tan, Bill Corcoran, Xingyuan Xu, Andrew Boes, Jiayang Wu, Thach Nguyen, S.T. Chu, B. E. Little, Roberto Morandotti, Arnan Mitchell, and David J. Moss, "Ultra-high bandwidth optical data transmission with a microcomb", IEEE Topical Meeting on Microwave Photonics (MPW), pp. 78-82.Virtual Conf., Matsue, Japan, November 24-26, 2020. Electronic ISBN:978-4-88552-331-1. **DOI:** 10.23919/MWP48676.2020.9314476

[108] M. Tan, X. Xu, J. Wu, R. Morandotti, A. Mitchell, and D. J. Moss, "RF and microwave high bandwidth signal processing based on Kerr Micro-combs", Advances in Physics X, VOL. 6, NO. 1, 1838946 (2020). DOI:10.1080/23746149.2020.1838946.

[109] Mengxi Tan, Xingyuan Xu, Jiayang Wu, Thach G. Nguyen, Sai T. Chu, Brent E. Little, Roberto Morandotti, Arnan Mitchell, and David J. Moss, "Photonic Radio Frequency Channelizers based on Kerr Micro-combs and Integrated Micro-ring Resonators", JOSarXiv.202010.0002.

[110] Mengxi Tan, Xingyuan Xu, David Moss "Tunable Broadband RF Photonic Fractional Hilbert Transformer Based on a Soliton Crystal Microcomb", Preprints, DOI: 10.20944/preprints202104.0162.v1

[111] Mengxi Tan, X. Xu, J. Wu, T. G. Nguyen, S. T. Chu, B. E. Little, R. Morandotti, A. Mitchell, and David J. Moss, "Orthogonally polarized Photonic Radio Frequency single sideband generation with integrated micro-ring resonators", *Journal of Semiconductors* **42** (4), 041305 (2021). DOI: 10.1088/1674-4926/42/4/041305.

[112] B. Corcoran, et al., "Ultra-dense optical data transmission over standard fiber with a single chip source", Nature Communications, vol. 11, Article:2568, 2020. DOI:10.1038/s41467-020-16265-x.

[113] X. Xu, et al., "Photonic perceptron based on a Kerr microcomb for scalable high speed optical neural networks", Laser and Photonics Reviews, vol. 14, no. 8, 2000070, 2020. DOI:10.1002/lpor.202000070.

[114] X. Xu, et al., "11 TOPs photonic convolutional accelerator for optical neural networks", Nature, vol.589 (7840) 44-51 (2021). DOI: 10.1038/s41586-020-03063-0.

[115] X Xu et al., "11 TeraFLOPs per second photonic convolutional accelerator for deep learning optical neural networks", arXiv preprint arXiv:2011.07393 (2020).

[116] D. Moss, "11 Tera-FLOP/s photonic convolutional accelerator and deep learning optical neural networks", Research Square (2021). DOI: https://doi.org/10.21203/rs.3.rs-493347/v1.

[117] D.Moss, "11.0 Tera-FLOP/second photonic convolutional accelerator for deep learning optical neural networks", TechRxiv. Preprint (2020). https://doi.org/10.36227/techrxiv.13238423.v1.

[118] Moss, David. "11 Tera-flop/s Photonic Convolutional Accelerator for Optical Neural Networks." OSF Preprints, 23 Feb. (2021). DOI: 10.31219/osf.io/vqt4s.